\documentclass[12pt]{article}
\usepackage{psfig}
\usepackage{amsmath}
\begin{document}

 \title{Polarization of the Tagged Compton Backscattered Laser Photons.\\
 Results of Monte Carlo Simulation. }
 \author{\sl A.~S.~Omelaenko, Yu.~P.~Peresun'ko,
\footnote{E-mail:peresunko@kipt.kharkov.ua} \\
\sl Yu.~M.~Ranyuk\footnote{E-mail:ranyuk@kipt.kharkov.ua}
, I.~M.~Shapoval,\\
National Science Center\\ Kharkov Institute of Physics and Technology,\\
 310108,Kharkov, Ukraine,\\
 \sl and V.~G.~Nedorezov\footnote{E-mail:nedorezov@aviva.inr.ac.ru}.\\
 Kurchatov Institute,123182 , Moscow.}
 \date{}
 \maketitle

 \begin{abstract}
 Polarization characteristics of the gamma beam  obtained by the
 Compton back scattering of laser photons on high energy electrons are
 evaluated by Monte-Carlo simulations.
 It is assumed that outgoing photons are tagged;  the energy dispersion of
 the tagging photons and emittance of the initial electrons are
 taken into account. Dependence of the final photon polarization parameters
 on measured photon energy is obtained. It is shown that polarization of
 final photons is decreasing with change for the worth of the tagging energy
 resolution. Calculations have been applied for the storage ring SIBERIA-2
 at Kurchatov Institute.  The obtained results indicate a  reasonability
 for construction of gamma-polarimeters on existing and planned facilities
 for the on-line measurement of the final photon beam polarization parameters.
 \end{abstract}

 \section{Theoretical description of the process of Compton scattering}
 Method of obtaining monochromatic and polarized photon beams of high energy
 and intensity by the Compton back scattering of laser photon on high-energy
 electron beam is widely used in many active and planned facilities
  \cite{1,2}. Idea of using the Compton back scattering of laser photons on
 relativistic electron beams as source of
 monochromatic and polarized gamma - radiation was specified in works of
 Harutyunian and Tumanian\cite{3,4} and in work of Milborn\cite{5},
 although polarization phenomenons in this reaction were considered
 earlier\cite{6}.
 Authors\cite{3,4} used in their evaluations  formula for Stoke's
 parameters of final photon from monograph\cite{7} which contained
 unfaithful expression for Stoke's parameter $\xi _2^{(f)}$
 of final photon circular polarization.
 In publishing 1969 of this monograph without commentaries was brought
 correct expression for this value. However, on the strength of
 popularity of works \cite{3,4}, casus with parameter $\xi _2^{(f)}$
 is discussed recently\cite{8}.
 \\ Important element in practical using of considered process
is a tagging system, which allows to measure an energy of outgoing photons.
Influence of this system on polarization features of final photon beam
 hitherto not explored sufficiently.
 The purpose of this work is the detail analysis of the polarization
 characteristics of the outgoing photon beam and studying of influence on it
 of the different parameters - the electron beam divergence, the
 uncertainties in definition of the outgoing photon energy by tagging of the
 final electrons and spin characteristics of initial electron beam.

\begin{figure}[tbh]
 \centerline{ \psfig{figure=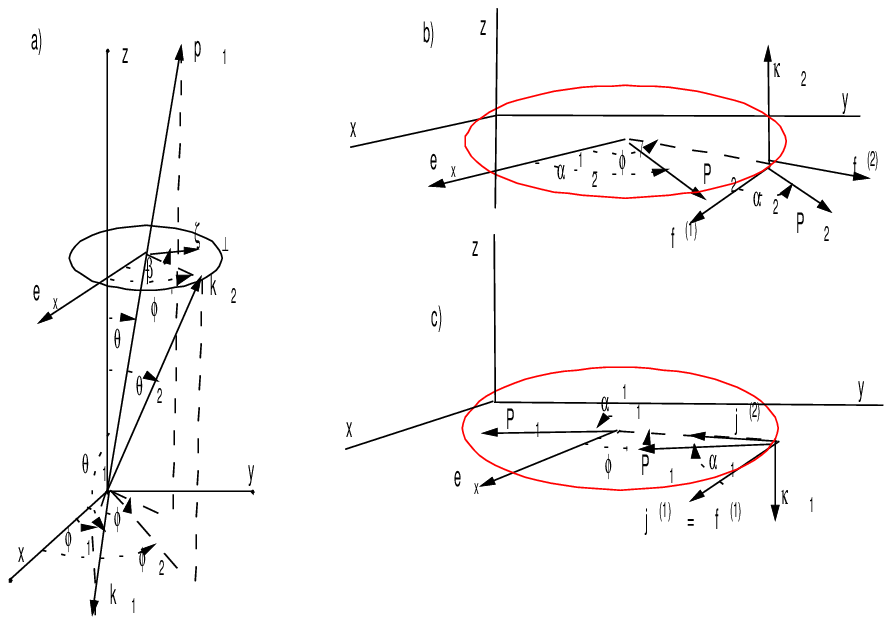,height=8.0 cm}}
\caption{}
 \label{fig.1}
\end{figure}

 Let's consider the kinematics of the process, that is shown in
  Figure~\ref{fig.1}.
 In the laboratory frame where axis $z$ coincides with a beam's axis and $x$
 axis belongs to horizontal plane of accelerator's beam initial electron with
 a momentum $\vec{p}_1$ ( polar angle --- $\theta $, azimuthal angle --- $
 \varphi $) interacts with a laser photon with a momentum $\overrightarrow{k}
 _1$ ( polar angle --- $\theta _1$, azimuthal angle --- $\varphi _1$).
 Scattered photon has the 3--momentum $\overrightarrow{k}_2$ ( polar angle
 --- $\theta _2$, azimuthal angle --- $\varphi _2$). Momentum of the final
 electron $\overrightarrow{p}_2$ is not shown. From 4--momentum conservation
 law $k_1+p_1=k_2+p_2$ follows $(k_2\cdot (p_1+k_1))=(k_1\cdot p_1)$, or
 \begin{eqnarray}
 &&\omega _2\{1-v\cdot [\cos \theta \cdot \cos \theta _2+\sin \theta \cdot
 \sin \theta _2\cos (\varphi _2-\varphi )]  \nonumber \\
 &&+\frac{\omega _1}{\varepsilon _1}\cdot [1-\cos \theta _1\cdot \cos \theta
 _2-\sin \theta _1\cdot \sin \theta _2\cdot \cos (\varphi _1-\varphi _2)]\}
 \label{eq.1} \\
 &=&\omega _1\cdot \{1-v\cdot [\cos \theta \cdot \cos \theta _1+\sin \theta
 \cdot \sin \theta _1\cdot \cos (\varphi _1-\varphi )]\}  \nonumber
 \end{eqnarray}
 where $\omega _1,\omega _2$ --- energies of the initial and final photons,
 $ \varepsilon _1$ --- energy of initial electron; $v=\sqrt{1-m^2/\varepsilon
 _1^2\ }$ --- velocity of initial electron. The angles $\theta $, $\varphi$
 of the initial electron are random values. Usually are considered values
 $\theta _x=\theta \cos \varphi $ and $\theta _y=\theta \sin \varphi $ that
 supposed to be normally distributed with a means $\theta _x^o=0$ and $\theta
 _y^o=0$ and variances $\sigma _x$ and $\sigma _y$ correspondingly. Electron
 beams of the modern storage rings have $\sigma _x\ \sim \ 10^{-3}\div
 10^{-4} $ rad, $\sigma _y$ $\ \sim \ 10^{-4}\div 10^{-5}$ rad. The range of
 allowed angles of final photon $\theta _2$, $\varphi _2$ is determined by
 the collimator's size and is of the order $10^{-3}-10^{-4}$~rad.

 So we can with a good accuracy change in (\ref{eq.1}):
 \[
 \sin \theta \cdot \cos \varphi \to \theta _x,\quad
 \sin \theta \cdot
 \sin \varphi \to \theta _y,
 \quad \sin \theta _2\to \theta _2,
 \]

 \[
 \cos \theta \to 1-\frac{1}{2}\theta ^2,
 \quad \cos \theta _2\to 1-
 \frac{1}{2}\theta _2^2.
 \]

 Replacing variables
 $$
 \left[ \theta _2,\varphi _2\right] \to \left[ \alpha _x=\theta _2\cos
 \varphi _2,\quad
 \alpha _y=\theta _2\sin \varphi _2\right] ,
 $$
 we can write (\ref{eq.1}) in the form:
 \begin{equation}
 \left( \alpha _x-\theta _x-\frac{\omega _1}{\varepsilon _1}
 \sin \theta _1\cos \varphi _1\right) ^2
 +\left( \alpha _y-\theta _y-\frac{\omega _1}
 {\varepsilon _1}\sin \theta \sin \varphi _1\right) ^2=r^2
 \label{eq.2}
 \end{equation}
 where
 \[
 r^2=\gamma ^{-2}\left[ x
 \left( \frac{\varepsilon _1-\omega _2}{\omega _2}
 \right) -1\right] ;\qquad
 \gamma =\frac{\varepsilon _1}{m};
 \qquad x=2
 \frac{\left( k_1\cdot p_1\right) }{m^2}
 \]

 Equation (\ref{eq.2}) is valid with an accuracy up to terms
 \[
 \left[ \frac{\omega _1}{\varepsilon _1},\quad \theta _2^2,
 \quad \theta _x^2,\quad
 \theta _y^2,\quad
 \gamma ^{-2}\right] \ll 1
 \]

 Thus we can see from (\ref{eq.2}) that photons with a definite energy
 $\omega _2$
 are emitted at the surface of the circle cone with an axes along
 direction of the vector
 $\vec{l}=\vec{n}+\frac{\omega _1}{\varepsilon _1}\vec{\kappa }_1$
 and the opening angle
 $r$
 $(\vec{n}=\vec{p}_1/\left| \vec{p}_1\right|$,
 $\vec{\kappa }_1=\vec{k}_1/\left| \vec{k}_1\right| )$

  From the condition of positive definition of $r^2$ maximal possible energy
 of secondary photon is
 \begin{equation}
 \omega _{2\max }=\varepsilon _1\frac{x}{1+x}  \label{eq.3}
 \end{equation}

 An invariant form of the cross section of the Compton scattering of the
 polarized photon by the polarized electron when\ one detect final photon
 with a Stoke's parameters
 $\widetilde{\xi }_i^{\left( f\right) }$
 is \cite{7,10}:
 \begin{equation}
 \frac{d^2\sigma }{dyd\varphi _2}=
 Sp\{ \widehat{\rho }_c\cdot \widehat{\rho }^{(f)} \}
 =\frac{r_o^2}{2x^2}\left\{ F_0+\widetilde{\xi }_1^{\left( f\right)}
 F_1+\widetilde{\xi }_2^{\left( f\right) }F_2+\widetilde{\xi }_3^{\left(
 f\right)}F_3\right\} ;  \label{eq.4}
 \end{equation}
 where
 \[
 \widehat{\rho }_c=\frac{r_o^2}{2x^2}\left(
\begin{tabular}{cc}
$F_0+F_3$ & $F_1+iF_2$ \\
$F_1-iF_2$ & $F_0-F_3$
\end{tabular}
 \right) ;
 \]

 \[
 \widehat{\rho }^{(f)}=\frac 12\left(
 \begin{tabular}{cc}
 $1+\widetilde{\xi }_3^{\left( f\right) }$ & $\widetilde{\xi }_1^{\left(
 f\right) }+i\widetilde{\xi }_2^{\left( f\right) }$ \\
 $\widetilde{\xi }_1^{\left( f\right) }-i\widetilde{\xi }_2^{\left( f\right)
 } $ & $1-\widetilde{\xi }_3^{\left( f\right) }$
 \end{tabular}
 \right) ;
 \]
 \[
 F_0=\frac xy+4z\left( 1+z\right) \left( 1-\widetilde{\xi }_3^{\left(
 1\right) }\right) -2\widetilde{\xi }_2^{\left( 1\right) }\ z\left[ \left(
 1+2\ z\right) \left( s\cdot k_1\right) +(s\cdot k_2\right)] ;
 \]
 \[
 F_1=2\widetilde{\xi }_1^{\left( 1\right) }\ \left( 1+2z\right)
 -4\widetilde{
 \xi }_2^{\left( 1\right) }\frac zy\left( sk_1p_1k_2\right) ;
 \]
 \begin{eqnarray*}
 F_2 &=&\widetilde{\xi }_2^{(1)}(1+2z)(\frac xy+\frac yx)-2z[(s\cdot
 k_1)+(1+2z)(s\cdot k_2)] \\
 &&+2\widetilde{\xi }_3^{(1)}z[(1+2z)(s\cdot k_2)-\frac yx(s\cdot k_1)]+4
 \widetilde{\xi }_1^{(1)}\frac zx(sk_1p_1k_2);
 \end{eqnarray*}
 \[
 F_3=-4z1+z\left( 1-\widetilde{\xi }_3^{\left( 1\right) }\right)
 +2\widetilde{
 \xi }_3^{\left( 1\right) }+2\widetilde{\xi }_2^{\left( 1\right) }z[\left(
 1+2z\right) \left( s\cdot k_1\right) -\frac xy\left( s\cdot k_2\right) ];
 \]
 \[
 x=2\frac{\left( k_1\cdot p_1\right) }{m^2},\quad y=2\frac{\left( k_2\cdot
 p_1\right) }{m^2},\quad z=\frac 1x-\frac 1y;
 \]

 $s_\mu $ --- 4--vector of the initial electron polarization,

 $(sk_1p_1k_2)=\varepsilon ^{\mu \nu \rho \sigma }s_\mu k_{1\nu }p_{1\rho
 }k_{2\sigma }/m^3,\quad \varepsilon ^{0123}=+1.$

 $r_o$---classical radius of electron, $\varphi _2$--azimuthal angle of the
 final photon (see figure \ref{fig.1}).

 Here Stoke's parameters of the initial photon $\widetilde{\xi }_i^{(1)}$ are
 defined relatively to the relativistic invariant unit vectors \cite{9,10}:
 \begin{equation}
 j_\mu ^{(1)}=\varepsilon _{\mu \nu \rho \sigma }k_{1\nu }p_{1\rho
 }k_{2\sigma }/a,\qquad j_\mu ^{(2)}=\frac 2{m^2x}\varepsilon _{\mu \nu \rho
 \sigma }k_{1\nu }p_{1\rho }j_\sigma ^{(1)};  \label{eq.5}
 \end{equation}
 where

 $a=\frac{m^3}2\left[ \left( x-y\right) \left( xy+y-x\right) \right] ^{1/2}$,

 and Stoke's parameters of the final photon, $\widetilde{\xi}_i^{(f)}$,- to
 the unit vectors:
 \begin{equation}
 f_\mu ^{(1)}=j_\mu ^{(1)};\qquad f_\mu ^{(2)}=2/\left( m^2y\right)
 \varepsilon _{\mu \nu \rho \sigma }k_{2\nu }p_{1\rho }f_\sigma ^{(1)}
 \label{eq.6}
 \end{equation}
 we shall use following parametrization of the photon's polarization
 characteristics \cite{11}:
 \begin{equation}
 \xi _1=P\cdot \cos 2\beta \cdot \sin 2\alpha ;\qquad \xi _3=P\cdot \cos
 2\beta \cdot \cos 2\alpha ;\qquad \xi _2=P\cdot \sin 2\beta  \label{eq.7}
 \end{equation}
 where $P=\sqrt{\xi _1^2+\xi _2^2+\xi _3^2}$--degree of the total
 polarization,

 $P_l=\sqrt{\xi _1^2+\xi _3^2}$--degree of the linear photon polarization,

 $\alpha \ $--the angle between direction of axis $\vec{j}^{(1)}$ and
 direction of maximal linear polarization of photon that is reading counter
 clockwise when one looks from the end of photon momentum vector.

 Gauge transformation permits to choose unit vectors $j_\mu ^{(i)}$, $f_\mu
 ^{(i)}$ as a pure space vectors that are orthogonal to a $\vec{k_1}$ and
 $\vec{k_2}$ correspondingly.

 Such a vectors, $\ \vec{j}_{\perp }^{(1)}$, $\vec{j}_{\perp }^{(2)}$
 and $\vec{f}_{\perp }^{(1)}$, $\vec{f}_{\perp }^{(2)}$ are calculated
 in Appendix
 for the case of arbitrary angles of initial photon $\theta _1$, $\varphi _1$
 with an accuracy up to a small terms of the order of $\theta ^2$, $\theta
 _2^2$, $\gamma ^{-2}$ and have a form:
 \begin{eqnarray}
 \vec{j}_{\perp }^{(1)} &=&\vec{e}_x\left[ \sin \varphi ^{\prime }+\left(
 1+\cos \theta _1\right) \cos \varphi _1\sin \left( \varphi _1-\varphi
 ^{\prime }\right) \right]  \label{eq.8} \\
 &&+\vec{e}_y\left[ -\cos \varphi ^{\prime }+\left( 1+\cos \theta _1\right)
 \sin \varphi _1\sin \left( \varphi _1-\varphi ^{\prime }\right) \right]
 \nonumber \\
 &&-\vec{e}_z\sin \theta _1\sin \left( \varphi _1-\varphi ^{\prime }\right) ;
 \nonumber
 \end{eqnarray}
 \begin{eqnarray}
 \vec{j}_{\perp }^{(2)} &=&\vec{e}_x\left[ \cos \theta _1\cos \varphi
 _1+\left( 1+\cos \theta _1\right) \Pr \sin \varphi _1\sin \left( \varphi
 _1-\varphi ^{\prime }\right) \right]  \label{eq.9} \\
 &&+\vec{e}_y\left[ \cos \theta _1\sin \varphi ^{\prime }+(1+\cos \theta
 _1)\cos \varphi _1\sin \left( \varphi _1-\varphi ^{\prime }\right) \right]
 \nonumber \\
 &&-\vec{e}_z\sin \theta _1\cos \left( \varphi _1-\varphi ^{\prime }\right) ;
 \nonumber
 \end{eqnarray}
 \begin{equation}
 \vec{f}_{\perp }^{(1)}=\vec{e}_x\sin \varphi ^{\prime }-\vec{e}_y\cos
 \varphi ^{\prime };  \label{eq.10}
 \end{equation}
 \begin{equation}
 \vec{f}_{\perp }^{(2)}=\vec{e}_x\sin \varphi ^{\prime }+\vec{e}_y\cos
 \varphi ^{\prime };  \label{eq.11}
 \end{equation}

 Here $\varphi ^{\prime }$ is the azimuthal angle on the surface of the cone
 of the final photon emitting. This angle is reading from $\vec{e_x}\ $when
 direction of $\vec{p_1}$ serves as a polar axis (see figure~\ref{fig.1}).
 Expressions (\ref{eq.8} - \ref{eq.11}) are not valid, of course, in a very
 narrow interval of final photon energy $\omega _2$ near maximal final photon
 energy $\omega _{2\max }$ when defined in equation (\ref{eq.2}) value\\
 $r=\gamma ^{-1}[(1-x)(\omega _{2\max }-\omega _2)/\omega _2]^{1/2}\ $
 becomes less or order of $\{\gamma ^{-2},\ \theta ^2,\ \theta _2^2\}$. In
 such interval of energies of final photon $\omega _2$ linear polarization of
 outgoing photon may be strongly correlated with directions of initial photon
 and electron moments, but we shall not discuss this point in the current
 work.

 On the experiment polarization of photons is determined relatively to the
 fixed axes $\left\{ \vec{e_x},\ \vec{e_y},\ \vec{e_z}\right\} $ of the
 laboratory frame. In this frame Stoke's parameters of photons are $\xi
 _i^{(1)},\ \xi _i^{(f)}$ and can be expressed in terms of
 $\widetilde{\xi }_i^{(1)}$, $\widetilde{\xi }_i^{(f)}$ by the following way:

 Let accordingly to parametrization (\ref{eq.7}) vector of linear
 polarization of outgoing photon is $\vec{P_2}$ and has an angle $\alpha _2$
 with an axes $\vec{f}_{\perp }^{(i)}$ (see Fig.~\ref{fig.1}). Then it's
 angle $\widetilde{\alpha }_2\ $ with an axes $\vec{e_x}\ $ is
 $\widetilde{\alpha }_2=\alpha _2+(\varphi ^{\prime }-\frac \pi 2)$
 and we can write:
 \begin{eqnarray}
 \xi _1^{(f)} &=&P_2\cos 2\beta _2\sin 2\widetilde{\alpha }_2=P_2\cos 2\beta
 _2\sin 2(\alpha _2+\varphi ^{\prime }-\frac \pi 2)  \nonumber  \label{eq.666}
 \\
 &=&-\widetilde{\xi }_1^{(f)}\cos 2\varphi ^{\prime }-\widetilde{\xi }
 _3^{(f)}\sin 2\varphi ^{\prime };  \label{eq.12}
 \end{eqnarray}
 \begin{equation}
 \xi _3^{(f)}=-\widetilde{\xi }_3^{(f)}\cos 2\varphi ^{\prime }+\widetilde{
 \xi }_1^{(f)}\sin 2\varphi ^{\prime };  \label{eq.13}
 \end{equation}
 \begin{equation}
 \xi _2^{(f)}=\widetilde{\xi }_2^{(f)}  \label{eq.14}
 \end{equation}

 And for initial photon $\widetilde{\alpha _1}=\alpha _1-(\varphi ^{\prime
 }-\frac \pi 2)$ (angles $\alpha _1$ and $\varphi ^{\prime }$ are reading in
 opposite directions), therefore\footnote{%
 In this point work \cite{9} contains mistake.Besides that in expression
 for cros section in \cite{9} is missing factor $1/2$.
  Other formula of \cite{9}
 which we used have been checked and are correct.}:
 \begin{equation}
 \xi _1^{(1)}=-\widetilde{\xi }_1^{(1)}\cos 2\varphi ^{\prime }+
 \widetilde{
 \xi }_3^{(1)}\sin 2\varphi ^{\prime };  \label{eq.15}
 \end{equation}
 \begin{equation}
 \xi _3^{(1)}=-\widetilde{\xi }_3^{(1)}\cos 2\varphi ^{\prime }
 -\widetilde{
 \xi }_1^{(1)}\sin 2\varphi ^{\prime };  \label{eq.16}
 \end{equation}
 \begin{equation}
 \widetilde{\xi }_2^{(1)}=\xi _2^{(1)}.  \label{eq.17}
 \end{equation}

 Here and below we put $\theta _1=\pi $. It is easy to see that uncertainties
 in determination of $\ \theta _1\ $do not affect on polarization
 characteristics of outgoing photon up to terms of order $(\Delta \theta
 _1)^2$.

 Thus one can see that natural and convenient variables for describing of
 considered process of Compton backscattering of laser photon on an
 relativistic electron with a tagging of final photons when on experiment is
 straightforward determined the energy of this photon $\omega _{2}$%
 are variables $u=\gamma r$ and $\varphi ^{\prime }.$In this variables after
 substitution of (\ref{eq.12}--\ref{eq.17}) considered cross section has a
 form (see \cite{9}):
 \begin{equation}
 \frac{d^2\sigma }{d\omega _2\ d\varphi ^{\prime }}=\frac{r_o^2}{2\varepsilon
 _1x\left( 1+u^2\right) ^2\left( 1+x+u^2\right) }\left\{ \Phi _0
 +\widetilde{
 \xi }_1^{(f)}\Phi _1+\widetilde{\xi }_2^{(f)}\Phi _2+\widetilde{\xi}
 _3^{(f)}\Phi _3\right\}  \label{eq.18}
 \end{equation}
 where
 \begin{eqnarray}
 \Phi _0 &=&2+2x+x^2+u^2\left( 2+x^2\right) +2u^4\left( 1+x\right) +2u^6
 \nonumber \\
 &&-4\left( \xi _3^{(1)}\cos 2\varphi ^{\prime }-\xi _1^{(1)}\sin 2\varphi
 ^{\prime }\right) u^2\left( 1+x+u^2\right)  \label{eq.19} \\
 &&-\xi _2^{(1)}[\lambda x\left( 1-u^2\right) \left( 2+x+2u^2\right) +2\zeta
 _{\perp }\cos \left( \beta -\varphi ^{\prime }\right) xu\left( 1+u^2\right)
 ];  \nonumber
 \end{eqnarray}
 \begin{eqnarray}
 \Phi _1 &=&2(1+x+u^2)(\xi _1^{(1)}(-1+u^4\cos 4\varphi ^{^{\prime }})+u^4\xi
 _3^{(1)}\sin 4\varphi ^{^{\prime }}  \nonumber \\
 &&-2u^2\sin 2\varphi ^{^{\prime }}+\xi _2^{(1)}\varsigma _{\bot }xu\sin
 (3\varphi ^{^{\prime }}-\beta ));  \label{eq.20}
 \end{eqnarray}
 \begin{eqnarray}
 \Phi _2 &=&-\xi _2^{(1)}\left( 1-u^2\right) (2+2x+x^2+2u^2\left(
 2+x+u^2\right) )  \nonumber \\
 &&+\lambda x\left( 2+x+xu^2+2u^4\right) +4\left( -\xi _3^{(1)}\cos 2\varphi
 ^{\prime }+\xi _1^{(1)}\sin 2\varphi ^{\prime }\right) \lambda xu^2
 \nonumber \\
 &&+2\left( 1+\xi _3^{(1)}\cos 2\varphi ^{\prime }-\xi _1^{(1)}\sin 2\varphi
 ^{\prime }\right) \zeta _{\perp }xu\left( 1-u^2\right) \cos \left( \beta
 -\varphi ^{\prime }\right)  \nonumber \\
 &&+2\left( \xi _1^{(1)}\cos 2\varphi ^{\prime }+\xi _3^{(1)}\sin 2\varphi
 ^{\prime }\right) \zeta _{\perp }\sin \left( \beta -\varphi ^{\prime
 }\right) xu\left( 1+u^2\right) ;  \label{eq.21}
 \end{eqnarray}
 \begin{eqnarray}
 \Phi _3 &=&2\left( 1+x+u^2\right) [\xi _3^{(1)}(1+u^4\cos 4\varphi ^{\prime
 })-\xi _1^{(1)}u^4\sin 4\varphi ^{^{\prime }}  \nonumber \\
 &&-2u^2\cos 2\varphi ^{^{\prime }}+\xi _2^{(1)}\zeta _{\perp }xu\cos \left(
 \beta -\varphi ^{\prime }\right) ];  \label{eq.22}
 \end{eqnarray}

 Here
 \[
 u^2=\left( \gamma r\right) ^2=\left( 1+x\right) \frac{\omega _{2\max
 }-\omega _2}{\omega _2};
 \]
 $\lambda $--longitudinal polarization of initial electron, $\xi _{\perp }$
 -transverses polarization of initial electron, $\beta $- angle between x
 axes and direction of transverses polarization of electron.

 According to general theory from (\ref{eq.18}-\ref{eq.22}) follows that
 proper Stoke's parameters of outgoing photon determined relatively to the
 laboratory frame axes $\left\{ x,y,z\right\} $ are:
 \begin{equation}
 \xi _1^{(2)}=\frac{\Phi _1}{\Phi _0},\ \xi _2^{(2)}=\frac{\Phi _2}{\Phi _0}
 ,\ \xi _3^{(2)}=\frac{\Phi _3}{\Phi _0}  \label{eq.23}
 \end{equation}

 It would be mentioned that in realistic experiment of Compton backscattering
 of laser photons when initial relativistic electrons have some angle
 dispersion there is not existing one--to--one correspondence between
 direction of emitting of final photon and it's polarization characteristics.
 Actually, photon with defined energy $\omega _2$ and angles $\theta
 _2,\varphi _2$(see figure~\ref{fig.1}) may be emitted by the whole set of
 initial electrons that have direction of motion situated at the surface of
 cone with corresponding to $\omega _2$ angle of opening $r$.

 Corresponding to each such event angle $\varphi ^{^{\prime }}$ is different
 and therefore will be different polarization of emitted photon. Hence, we
 have to consider only averaged over all possible events of Compton
 interaction polarization characteristics of outgoing photon. We should like
 to emphasize that for calculation of different averaging polarization
 parameters (for example $\left\langle \xi _i^{\left( 2\right)}\right\rangle
 $, one has to average values $\Phi _i$ and to build $\left\langle \xi
 _i^{\left( 2\right) }\right\rangle $ by using values $\left\langle \Phi
 _i\right\rangle $.

 \section{Some details of calculations.}

 It follows from the previous consideration that the event generator for the
 Monte-Carlo simulation of the process of Compton backscattering of laser
 photon by accelerated electron beam when outgoing photon beam is tagged and
 collimated has to consist of following parts:

 \begin{enumerate}
 \item  Generator of random numbers for the values $\omega _2,\ \varphi
 ^{^{\prime }}$ with appropriate distribution;

 \item  Although values which describe direction of initial electron motion,
 do not enter to expressions (\ref{eq.18}) it could effect on distribution of
 $\omega _2,\ \varphi ^{^{\prime }}$ through the condition of final photon
 passing across the collimator with radius $R$ situated at distance $L$ from
 the Compton interaction point:
 \begin{equation}
 \left( \theta _x+r\cos \varphi ^{^{\prime }}\right) ^2+\left( \theta
 _y+r\sin \varphi ^{^{\prime }}\right) ^2\le (R/L)^2  \label{eq.24}
 \end{equation}
 \end{enumerate}

 So we have to include appropriate generator for this values. Distribution of
 this values is well known -- with a good accuracy it described by the normal
 distribution with mean $\theta _x^{(0)}=\theta _y^{(0)}=0$ and variance
 $\sigma _x^2,\sigma _y^2$ correspondingly; $\varphi ^{^{\prime }}$ -
 uniformly distributed in the range $[0,2\pi ]$, and question about
 distribution of $\omega _2$ requests some discussion. As it is known, the
 procedure of tagging of the photons in the process of Compton backscattering
 of laser photons on electron beam consists of measuring of energy of the
 scattered electrons $\varepsilon _2$ that is connected with the energy of
 other taking part in this process particles by the simple equation
 $\varepsilon _1+\omega _1=\varepsilon _2+\omega _2$. Energy $\varepsilon _1$
 of initial electron and energy $\omega_1$ of laser photon is known with a
 high accuracy. Energy $\varepsilon _2$ (and therefore - $\omega _2$) usually
 is measured by the magnet spectrometer with some error. Otherwise speaking,
 $\omega _2$ is random value that is distributed with some appropriate
 distribution and result of it's measuring - value of mean $\omega _{20}$ of
 this distribution. Distribution of this value has to be calculated by
 method of Monte-Carlo simulation for the cases of correspondent facilities.
 \ In our calculations should be used such a distributions for the
 value $\omega _2 $. But in the calculations of current article we restrict
 ourselves by using of the first two known moments of such distributions --
 the mean $\omega _{20}$ and variance $\sigma _\omega ^2$ and have
 approximated distribution of $\omega _2$ by the modified Gaussian
 distribution:
 \begin{equation} f(\omega _2)=\frac A{\sigma _\omega
 \sqrt{2\pi }}\exp \{-\frac{(\omega _2-\omega _{20})^2}{2\sigma _\omega
 ^2}\}\Theta (\omega _2)\Theta (\omega _{2\max }-\omega _2)  \label{eq.25}
 \end{equation}

 Where $A$ - an appropriate normalization factor.

 That is events that did not belong to the bounded physical region $0\leq
 \omega _2\leq \omega _{2\max }$ has been rejected. Thus, in our calculations
 we have used an event generator that consists of generator of the set of
 random numbers \{$\theta _x$,$\theta _y$,$\omega _2$,$\varphi ^{^{\prime }}$
 \} with normal distributions for \{$\theta _x$,$\theta _y$\}, uniformly
 distributed $\varphi ^{^{\prime }}$,and distribution (\ref{eq.25} for
 $\omega _2$ with some defined value of mean $\omega _{29}$,
 Each generated
 set checked by the condition \ref{eq.24}, and then calculated values $\Phi
 _i^{(n)}$ for obtained values of \{$\omega _2^{(n)}$, $\varphi ^{^{\prime
 }(n)}$\}.

 Averaged values have been calculated by the formula:
 \begin{equation}
 \left\langle \Phi _i\right\rangle =\frac 1N\sum\limits_{n=1}^N\Phi _i^{(n)}
 \label{eq.26}
 \end{equation}

 and it's associated error as follows:
 \begin{equation}
 \bigtriangleup \Phi _i=\frac 1{\sqrt{N(N-1)}}\left\{
 \sum\limits_{n=1}^N\left( \Phi _i^{(n)}\right) ^2-N\left\langle \Phi
 _i\right\rangle ^2\right\} ^{1/2}  \label{eq.27}
 \end{equation}
 Here $N$ --number of generated events.

 Averaged Stoke's parameters are defined as $\left\langle \xi
 _i^{(2)}\right\rangle =\left\langle \Phi _i\right\rangle /\left\langle \Phi
 _0\right\rangle $ and it's error could be estimated as follows:

  From (\ref{eq.27}) one can see that with probability $P$ confidence interval
 for $\Phi _i$ is:
 \[
 \left\langle \Phi _i\right\rangle -t_P\bigtriangleup \Phi _i\le \Phi _i\le
 \left\langle \Phi _i\right\rangle +t_P\bigtriangleup \Phi _i
 \]

 Here $t_P$ corespondent factor (we have used $P=.95$ and $t_P=1.96$).

 Therefore for ratio $\xi _i^{(2)}=\Phi _i/\Phi _0$ we will have
 corresponding confidence interval:
 \begin{equation}
 \left\langle \xi _i^{(2)}\right\rangle -\bigtriangleup \xi _i^{(2)}\le \xi
 _i^{(2)}\le \left\langle \xi _i^{(2)}\right\rangle +\bigtriangleup \xi
 _i^{(2)}  \label{eq.28}
 \end{equation}

 where $\bigtriangleup \xi _i^{(2)}=\frac{t_P}{\left\langle \Phi
 _0\right\rangle }\left\{ \bigtriangleup \Phi _i+\left| \left\langle \xi
 _i^{(2)}\right\rangle \right| \bigtriangleup \Phi _0\right\}$.

  From (\ref{eq.26}-\ref{eq.28}) one can see that by increasing of $N$ may be
 achieved any desired accuracy of calculation of $\left\langle \xi
 _i^{(2)}\right\rangle $.

 At the numerical calculations we have used following values of parameters,
 appropriated to SIBERIA-2 facility:
 \[
 \sigma _x=5.0\cdot 10^{-4},\quad \sigma _y=1.0\cdot 10^{-4},\quad \sigma
 _\omega =0.05\cdot \omega _{20},\quad L=18m,R=.02m
 \]
 \[
 \varepsilon _1=2.5GeV,\quad \omega _1=2.34eV
 \]

 \section{Results of calculations}

  From the beginning we should like to consider the case when initial laser
photon
 has pure circular polarization ($\xi _1^{(1)}=\xi _3^{(1)}=0;\xi _2^{(1)}=1.0
 $). Effects of initial electron's polarization will not be considered. From
 equation (\ref{eq.18} - \ref{eq.22}) one can see that in this case values
 $\Phi _i$ do not depend on $\varphi ^{^{\prime }}$ and only source of
 dispersion of cros section and polarization of outgoing photon is dispersion
 of measuring $\omega _2$. In the figure 2 are plotted values of
 outgoing photon averaged circular polarization $\left\langle \xi
 _2^{(2)}\right\rangle =\left\langle \Phi _2\right\rangle /\left\langle \Phi
 _0\right\rangle $ versus mean $\omega _{20}$ for the cases when variance
 $\sigma _\omega =0$ and $\sigma _\omega =0.05\ \omega _{20}$. Shown also error
 bar calculated by formula (\ref{eq.28}), $N=10^6$ for each $\omega _{20}$.
 In the case $\sigma _\omega =0$ final photon energy is measured exactly and
 $\omega _2=\omega _{20}$. Corresponding to this case, the curve of polarization
 dependence on $\omega _{20}$ could be regarded as theoretical curve for
 dependence of final photon polarization on  energy $\omega _2$.  One can see
 that taking into account dispersion of final photon energy measuring $\sigma
 _\omega $ don't leads to the increasing of final photon polarization
 dispersion but causes some systematical deviation in the dependence of
 averaged Stoke's parameter on measured photon energy $\omega _{20}$. This
 deviation increases when $\omega _{20}$ tends to the maximal bound
 $\omega _{2max}$and attains value about $1.3\%$ for $\sigma _\omega
 =0.05\ \omega _{20}$.

 In figure \ref{fig.3} are shown results of analogous calculations for value
 $\left\langle \xi _3^{(2)}\right\rangle $ for the case when initial photon is
 linear polarized, $\xi _3^{(1)}=1,\xi _1^{(1)}=\xi _2^{(1)}=0$. Averaging
 and estimation of errors has been provided with help of equations (\ref
 {eq.28}) for $N=10^6$ per each $\omega _{20}$. One can see that effect of
 deviation of dependence of final photon Stoke's parameters on $\omega _{20}$
 from theoretical dependence on $\omega _2$ (case $\sigma _\omega =0$) is
 observed for the case of linear polarization too. If $\sigma _\omega =0.05
 \omega _{20}$ then maximal deviation attains about $0.7\%$., and if $\sigma
 _\omega =0.1\ \omega _{20}$ , then maximal deviation is $\sim 3.3\%.$

 Mentioned effect points out the necessity of gamma--polarimeter to check
 the final photon beam polarization parameters at created and existing
 facilities .

 \section*{Appendix}

 Vector $j_\mu ^{(1)}$is a unit space-like vector of the form (\ref{eq.5}):
 \begin{equation}
 j_\mu ^{(1)}=\varepsilon ^{\mu \nu \rho \sigma }k_{1\nu }p_{1\rho
 }k_{2\sigma }/a;\qquad \left( j_\mu ^{(1)}\right) ^2=-1.  \tag{A.1}
 \end{equation}

 By the gauge transformation it may be done a pure space vector, orthogonal
 to $\overrightarrow{k}_1:$%
 \[
 \widetilde{j}_\mu ^{(1)}=j_\mu ^{(1)}+\beta _1k_{1\mu };\quad \widetilde{j}
 _0^{(1)}=);\widetilde{\overrightarrow{j}}^{(1)}=\overrightarrow{j}_{\perp
 }^{(1)}
 \]
 where
 \[
 \overrightarrow{j}_{\perp }^{(1)}=C_1\left[ \overrightarrow{j}^{(1)}-
 \overrightarrow{k}_1\left( \overrightarrow{j}^{(1)}\cdot \overrightarrow{k}
 _1\right) \right] ;\quad \left( \overrightarrow{j}_{\perp }^{(1)}\right)
 ^2=1,\quad \left( \overrightarrow{j}_{\perp }^{(1)}\cdot \overrightarrow{k}
 _1\right) =0,
 \]
 \[
 \left( \overrightarrow{j}^{(1)}\right) _m=C\left\{ \varepsilon
 ^{m0rs}k_{10}p_{1r}k_{2s}+\varepsilon ^{mn0s}k_{1n}p_{10}k_{2s}+\varepsilon
 ^{mnr0}k_{1n}p_{1r}k_{20}\right\} _m
 \]
 \begin{eqnarray*}
 &=&C\omega _1\varepsilon _1\omega _2\{-v\left[ \overrightarrow{n}\times
 \overrightarrow{\kappa }_2\right] +\left[ \overrightarrow{\kappa }_1\times
 \overrightarrow{\kappa }_2\right] -v\left[ \overrightarrow{\kappa }_1\times
 \overrightarrow{n}\right] \}_m \\
 &=&C\omega _1\omega _2\varepsilon _1\left[ \left( v\overrightarrow{n}-
 \overrightarrow{\kappa }_1\right) \times \left( v\overrightarrow{n}-
 \overrightarrow{\kappa }_2\right) \right] _m;
 \end{eqnarray*}
 \[
 v=\left| \overrightarrow{p}_1\right| /\varepsilon _1=\sqrt{1-\gamma ^{-2}}%
 ;\quad \overrightarrow{n}=\overrightarrow{p}_1/\left| \overrightarrow{p}%
 _1\right| ;\quad \overrightarrow{\kappa }_1=\overrightarrow{k}_1/\omega _1;\
 \overrightarrow{\kappa }_2=\overrightarrow{k}_2/\omega _2.
 \]

 That is
 \[
 \overrightarrow{j}_{\perp }^{(1)}=C_1\left( \left[ \left( v\overrightarrow{n}%
 -\overrightarrow{\kappa }_1\right) \times \left( v\overrightarrow{n}-%
 \overrightarrow{\kappa }_2\right) \right] -\overrightarrow{\kappa }_1\left(
 \overrightarrow{\kappa }_1\cdot \left[ \left( v\overrightarrow{n}-%
 \overrightarrow{\kappa }_1\right) \times \left( v\overrightarrow{n}-%
 \overrightarrow{\kappa }_2\right) \right] \right) \right)
 \]
 \begin{equation}
 =C_1\left( \left[ \left( v\overrightarrow{n}-\overrightarrow{\kappa }%
 _1\right) \times \left( v\overrightarrow{n}-\overrightarrow{\kappa }%
 _2\right) \right] -\overrightarrow{\kappa }_1v\left( \overrightarrow{n}[%
 \overrightarrow{\kappa }_1\times \overrightarrow{\kappa }_2]\right) \right)
 \tag{A 2.1}  \label{eq.A2.1}
 \end{equation}
 and from $\left| \overrightarrow{i}^{(1)}\right| ^2=1:$%
 \begin{eqnarray*}
 C_1^{-2} &=&1+v^2-\left( 1-v^2\right) \left( \overrightarrow{\kappa }_1\cdot
 \overrightarrow{\kappa }_2\right) ^2+2v\left( \overrightarrow{\kappa }%
 _1\cdot \overrightarrow{\kappa }_2\right) \cdot \\
 &&\left[ \left( \overrightarrow{\kappa }_1\cdot \overrightarrow{n}\right)
 +\left( \overrightarrow{\kappa }_2\cdot \overrightarrow{n}\right) -v\left(
 \overrightarrow{\kappa }_1\cdot \overrightarrow{n}\right) \left(
 \overrightarrow{\kappa }_2\cdot \overrightarrow{n}\right) -v\right] +
 \end{eqnarray*}
 \begin{equation}
 2v\left[ v\left( \overrightarrow{\kappa }_1\cdot \overrightarrow{n}\right)
 \left( \overrightarrow{\kappa }_2\cdot \overrightarrow{n}\right) -\left(
 \overrightarrow{\kappa }_1\cdot \overrightarrow{n}\right) -\left(
 \overrightarrow{\kappa }_2\cdot \overrightarrow{n}\right) \right]  \tag{A2.2}
 \label{eq.A2.2}
 \end{equation}

 Expanding (A2.1, A2.2) into a power series about $\theta
 _x,\theta _y,\alpha _x,\alpha _y$ (see equation\ref{eq.2}) we shall get:
 \[
 C_1^{-1}=\left( 1-\cos \theta _1\right) \sqrt{\left( \alpha _x-\theta
 _x\right) ^2+\left( \alpha _y-\theta _y\right) ^2}\left( 1+0\left( \gamma
 _{-2},\alpha _x,\alpha _y,\theta _x,\theta _y\right) \right)
 \]
 and with a same accuracy:
 \begin{eqnarray*}
 \overrightarrow{j}_{\perp }^{(1)} &=&\overrightarrow{e}_x\left[ \sin \left(
 \varphi ^{\prime }\right) +\frac{\sin ^2\theta _1}{1-\cos \theta _1}\cos
 \varphi _1\sin \left( \varphi _1-\varphi ^{\prime }\right) \right] \\
 &&+\overrightarrow{e}_y\left[ -\cos \varphi ^{\prime }+
 \frac{\sin ^2\theta _1
 }{1-\cos \theta _1}\sin \varphi _1\sin \left( \varphi _1-\varphi ^{\prime
 }\right) \right]
\end{eqnarray*}
 \begin{equation}
 -\overrightarrow{e}_z\sin \theta _1\sin \left( \varphi _1-\varphi ^{\prime
 }\right) ;  \tag{A3}  \label{eq.A3}
 \end{equation}
 where
 \[
 \sin \varphi ^{\prime }=\frac{\left( \alpha _y-\theta _y\right) }{\sqrt{%
 \left( \alpha _x-\theta _x\right) ^2+\left( \alpha _y+\theta _y\right) ^2}}%
 ;\quad \cos \varphi ^{\prime }=\frac{\left( \alpha _x-\theta _y\right) }{%
 \sqrt{\left( \alpha _x-\theta _x\right) ^2+\left( \alpha _y-\theta _y\right)
 ^2}}
 \]

 In the gauge when both $j_\mu ^{(1)}$ and $j_\mu ^{(2)}$are pure space
 vectors we have
 \begin{eqnarray*}
 \overrightarrow{j}_{\perp }^{(2)} &=&\overrightarrow{j}^{(2)}=\left[
 \overrightarrow{\kappa }_1\times \overrightarrow{j}_{\perp }^{(1)}\right] \\
 &=&\overrightarrow{e}_x\left[ \cos \theta _1\cos \varphi ^{\prime }+\left(
 1+\cos \theta _1\right) \sin \varphi _1\sin \left( \varphi _1-\varphi
 ^{\prime }\right) \right]
 \end{eqnarray*}
 \[
 +\overrightarrow{e}_y\left[ \cos \theta _1\sin \varphi ^{\prime }+\left(
 1+\cos \theta _1\right) \cos \varphi _1\sin \left( \varphi _1-\varphi
 ^{\prime }\right) \right]
 \]
 \begin{equation}
 -\overrightarrow{e}_z\sin \theta _1\cos \left( \varphi _1-\varphi ^{\prime
 }\right) .  \tag{A4}  \label{eq.A4}
 \end{equation}

 Analogous calculations for $f_\mu ^{(i)}$give
 \[
 \overrightarrow{f}_{\perp }^{(1)}=C_2\left\{ \left[ \left( v\overrightarrow{n%
 }-\overrightarrow{\kappa }_1\right) \times \left( v\overrightarrow{n}-%
 \overrightarrow{\kappa }_2\right) \right] -\overrightarrow{\kappa }_2v\left(
 \overrightarrow{n}\left[ \overrightarrow{\kappa }_1\times \overrightarrow{%
 \kappa }_2\right] \right) \right\} =
 \]
 \begin{equation}
 \overrightarrow{e}_x\sin \varphi ^{\prime }-\overrightarrow{e}_y\cos \varphi
 ^{\prime }+\overrightarrow{e}_z\cdot 0\left( \theta _1^2,\theta _2^2,\gamma
 ^{-2}\right)  \tag{A5}  \label{eq.A5}
 \end{equation}
 \begin{equation}
 \overrightarrow{f}_{\perp }^{(2)}=\left[ \overrightarrow{\kappa }_2\times
 \overrightarrow{f}_{\perp }^{(1)}\right] =\overrightarrow{e}_y\cos \varphi
 ^{\prime }+\overrightarrow{e}_x\sin \varphi ^{\prime }.  \tag{A6}
 \label{eq.A6}
 \end{equation}

 \begin{figure}[tbh]
  \begin{center}
    \psfig{figure=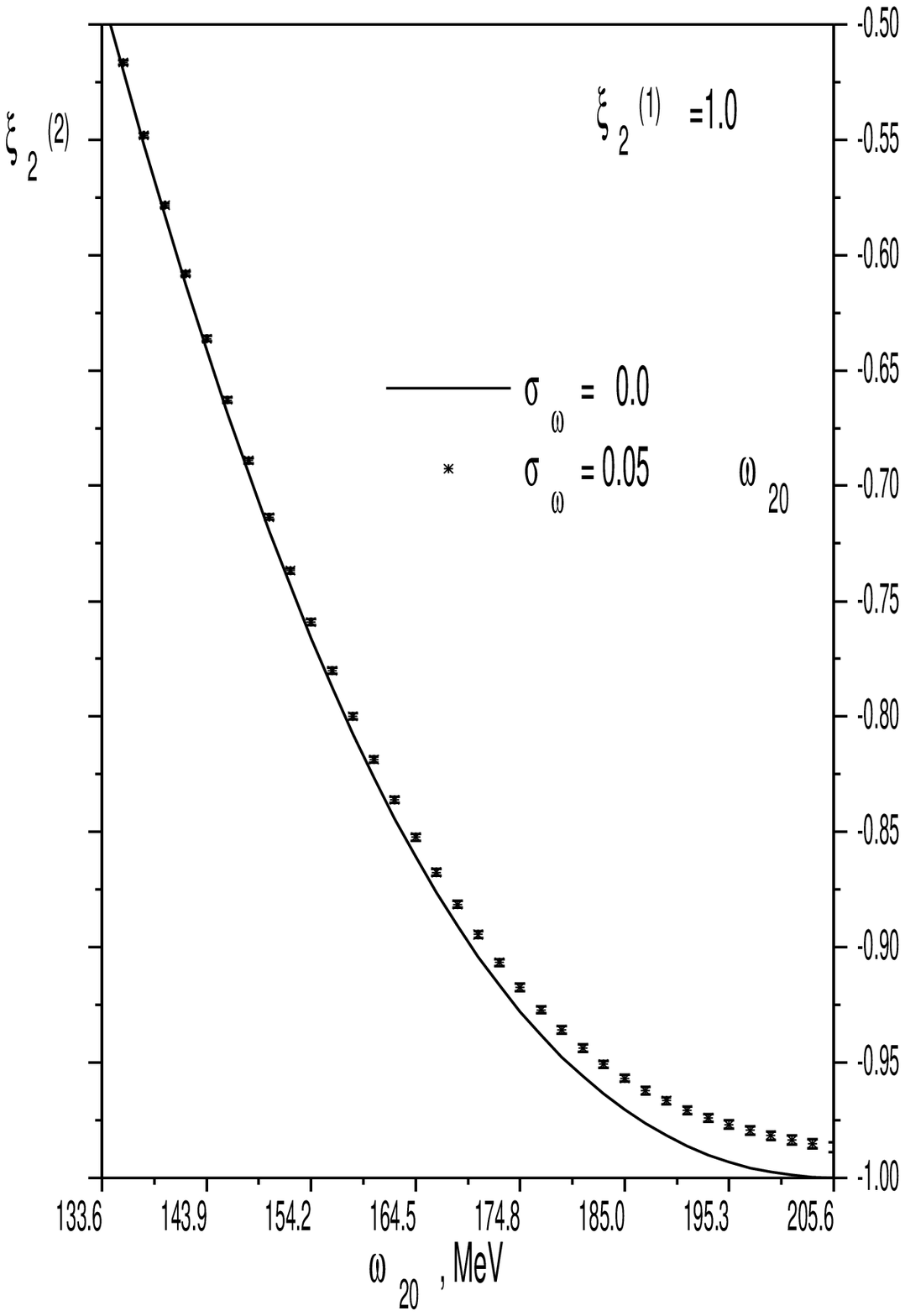,height=20cm}
   \caption{Averaged circular polarization of outgoing photon,\protect\\
$\left\langle \xi_3^{(2)}\right\rangle $ versus $\omega _{20}$.
 $\xi _2^{(1)}=0, \xi
_1^{(1)}=0,  \xi _3^{(1)}=1.0$.
 \protect\\ Number of simulated events -- $10^6$ for each value $\omega_{20}.$}
\end{center}
 \label{fig.2}
\end{figure}

\begin{figure}[tbh]
  \begin{center}
  \psfig{figure=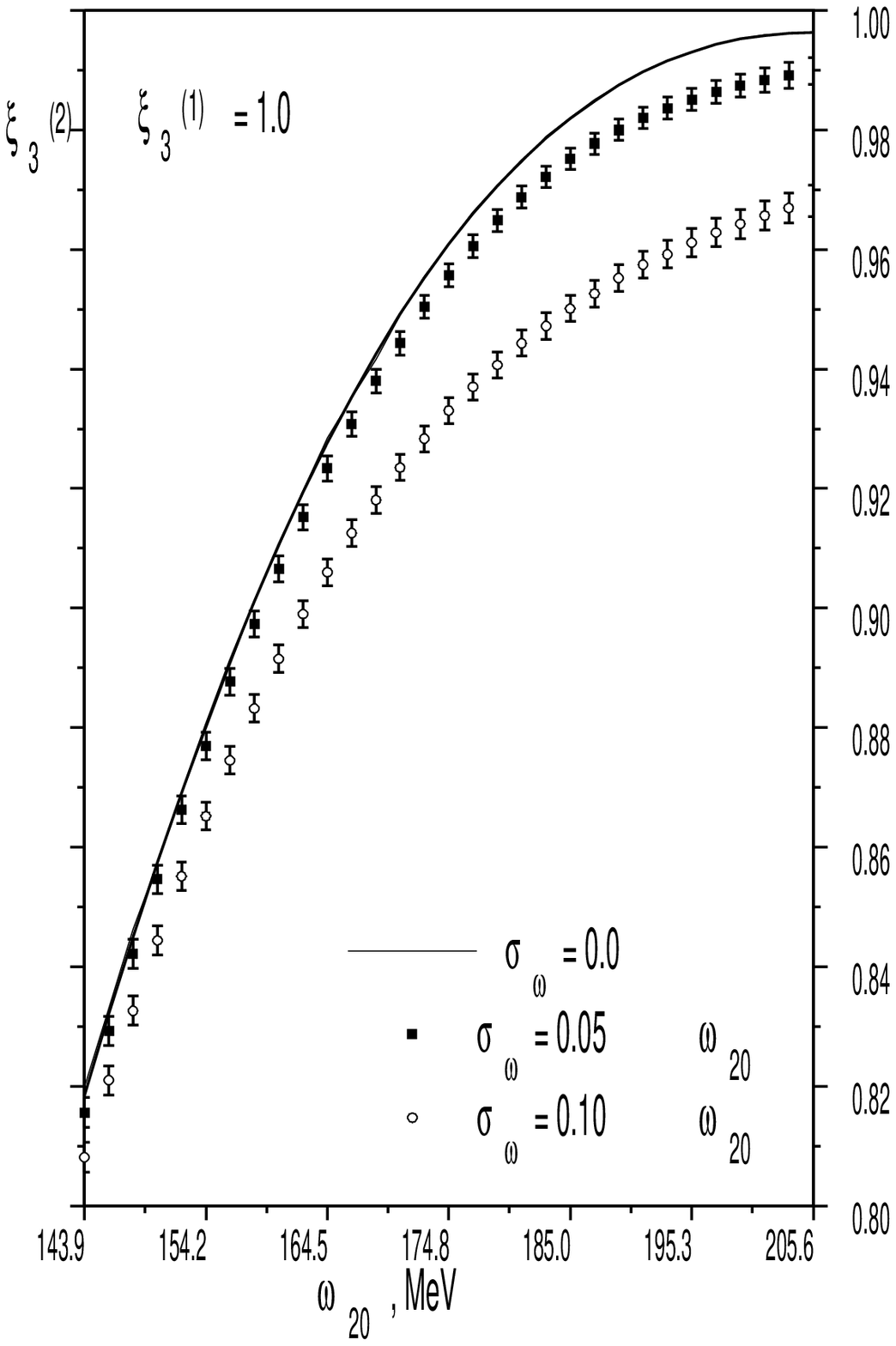,height=20cm}
 \caption{Averaged linear polarization of outgoing photon, \protect\\
 $\left\langle \xi _3^{(2)}\right\rangle $ versus $\omega
_{20}$. $\xi _2^{(1)}=0, \xi _1^{(1)}=0, \xi _3^{(1)}=1.0$.\protect\\
 Number of simulated events -- $10^6$ for each
value $\omega_{20}.$}
 \end{center}
 \label{fig.3}
\end{figure}

\end{document}